\documentclass[11pt,dvips,twoside,letterpaper]{article}

\usepackage{pslatex}
\usepackage{fancyhdr}
\usepackage{graphicx}
\usepackage{geometry}

\def\figurename{Figure} 
\makeatletter
\renewcommand{\fnum@figure}[1]{\figurename~\thefigure.}
\makeatother

\def\tablename{Table} 
\makeatletter
\renewcommand{\fnum@table}[1]{\tablename~\thetable.}
\makeatother

\usepackage{amsmath}
\usepackage{amssymb}
\usepackage{amsfonts}
\usepackage{amsthm,amscd}

\newtheorem{theorem}{Theorem}[section]

\theoremstyle{definition}

\newtheorem{remark}[theorem]{Remark}

\theoremstyle{remark}

\numberwithin{equation}{section}

\begin{document}
\begin{titlepage}

\begin{center}

\vspace{1cm}

\vskip .5in

{\Large\bfseries\scshape{Function expansion methods for solving autonomous nonlinear partial differential equations}}
\vspace{10pt}

 Mahouton Norbert Hounkonnou$^{\dag}$\footnote{Corresponding author:
norbert.hounkonnou@cipma.uac.bj  with copy to hounkonnou@yahoo.fr}
and Pascal Alain Dkengne Sielenou$^{\dag}$
\vspace{20pt}

{\em $^\dag$ International Chair in Mathematical Physics
and Applications} \\
{\em (ICMPA-UNESCO Chair)}\\
{\em  University of Abomey-Calavi}\\
{\em  072 B.P. 50 Cotonou, Republic of Benin}\\
\vspace{5pt}
E-mail: norbert.hounkonnou@cipma.uac.bj; sielenou$\_$alain@yahoo.fr

\today




\begin{abstract} \noindent
In this paper, we propose some  algorithms for analytical solution construction
 to nonlinear   polynomial partial differential equations with
constant  function coefficients. These schemes are based on
 one-(single), two-
 (double) or three- (triple) function expansion methods.
Most of the existing expansion function methods are well recovered
from the mentioned schemes. The effectiveness of these methods has been tested on some  nonlinear partial differential equations (NLPDEs)
describing important  phenomena in physics.
\end{abstract}
\end{center}

\noindent {\bf AMS Subject Classification:}  34A05, 34A34, 35A25

\vspace{.08in} \noindent \textbf{Keywords}: Nonlinear differential equations, function expansion methods, analytical solutions

\end{titlepage}
\makeatother

\section{Introduction}

A variety of  methods, e.g.
 the exponential function
\cite{16r}, the hyperbolic tangent function \cite{17r}, the Jacobi elliptic function expansion
 \cite{18r}, the sine-cosine function  \cite{19r},  the $F$-expansion  \cite{20r}, the projective
  Riccati equation expansion  \cite{21r}, the $\frac{G'}{G}$-expansion  \cite{22r},
  the $(\frac{G'}{G}, \frac{1}{G})$-expansion  \cite{23r} methods
  and their numerous extensions  \cite{29r,30r,37r,24r,33r,28r,25r,34r,26r,31r,32r,35r,36r,27r}, etc. to cite the main methods,
are used to obtain analytical solutions for nonlinear differential equations.
The common issue to all these methods is the rational form of the final solution.
The rationality here is with respect to either trigonometric, hyperbolic, elliptic, exponential functions, or  other given functions. Furthermore, the scope of
these methods is restricted to  polynomial differential equations.
We describe  in this paper  the general principle from which all the above mentioned methods derive
for the integration of autonomous nonlinear partial differential equations (NLPDEs). For the notations used in this work,
see  \cite{hk1} - \cite{hk5}.

\section{Function expansion methods: general procedure}

Consider the constant coefficient partial differential equations
\begin{equation}\label{3eqs1}
 F\left(u^{(s)}(x)\right)=0,
\end{equation}
where the nonzero positive integer $s$ is the order of the equation and $x=\left(x^1, \cdots  ,x^n\right)$ are independent variables.
The dependent variable $u=u(x)$ is a scalar valued function. If
\begin{equation}\label{3eqs2}
F\left(u^{(s)}(x)\right)=P\left( u_{(k_1)}[h_1](x), u_{(k_2)}[h_2](x) ,  \cdots   , u_{(k_r)}[h_r](x)  \right),
\end{equation}
where  $r,k_i,h_i\in \mathbb{N}$ with $\max\{ k_i, \,i=1, \cdots  ,r \}=s,$ $h_i\in \{  1,2,  \cdots , p_{k_i} \}$ and  $P$ is a polynomial  whose the indeterminates are the $r$ functions $u_{(k_i)}[h_i](x),$  the equation (\ref{3eqs1}) becomes
 \begin{equation}\label{3eqs3}
P\left( u_{(k_1)}[h_1](x), u_{(k_2)}[h_2](x) ,  \cdots   , u_{(k_r)}[h_r](x)  \right)=0
\end{equation}
 which is called a polynomial autonomous partial differential equation.

 The change of variables  $u(x)=v\left(\xi\right)$ with $\xi=\alpha_1 x^1+\alpha_2 x^2+  \cdots  +\alpha_n x^n,$ where $\alpha_i \in \mathbb{R}$ transforms
  (\ref{3eqs3}) into an ordinary differential equation
  \begin{equation}\label{3eqs4}
Q\left( v_{(k_1)}[1](\xi), v_{(k_2)}[1](\xi) ,  \cdots   , u_{(k_r)}[1](\xi)  \right)=0,
\end{equation}
where $Q$ is also a polynomial function whose the indeterminates are $v_{(k_i)}[1](\xi),$ $i=1, \cdots  ,r.$\\
If the polynomial $Q$ has only one monomial, then the equation (\ref{3eqs4}) is simply solved
by successive integrations. Thus, without loss of
generality, one can assume that the polynomial $Q$ is a sum of at least two monomials.
We propose in this paper to search for a solution of the equation (\ref{3eqs4}) by using
 single,   double  or  triple function expansion methods.

\subsection{Single function expansion method}

It consists to seek the function $v=v(\xi),$ solution of the equation (\ref{3eqs4}), into the form
\begin{equation}\label{3eqm1}
v=A(F)+F_{(1)}[1]B(F),
\end{equation}
where
$$ A(F)=\sum_{i=-m}^{m}a_iF^i,\quad   B(F)=\sum_{i=-\hat{m}}^{\hat{m}}b_iF^i   $$
and the function $F=F(\xi)$ is a solution to an auxiliary differential equation which is in one of the forms
 \begin{equation}\label{3eqm2}
F_{(1)}[1]= R   \quad \emph{\emph{or}} \quad \left( F_{(1)}[1]\right)^2= R,
\end{equation}
$R$ being a rational function in $F$ and $m, \hat{m}\in \mathbb{N},$ $a_i,b_i\in \mathbb{R}.$ The higher
order derivatives of $F$ can be written into  the form
$$
F_{(k)}[1]= R^{0,k}+ F_{(1)}[1]\,R^{1,k}  ,
$$
where the positive integer $k\geq 2$ and  $R^{i,k},$ $i=0, 1,$ are also rational functions in $F.$

One can adopt the following  iterative process  in practice for the determination of different parameters of  $v.$
\subsection*{Estimation of the integers $m, \hat{m}.$}

 Let $M_1, M_2,  \cdots  ,M_\nu$ be the monomials of $Q$ such that $M_1$ contains the highest order  derivative
of the function $v(\xi)$  and let $M_2$ be a nonlinear or linear monomial. $M_2$ is linear only in the case where all
remaining monomials are linear. Substitute in (\ref{3eqs4}) the expression (\ref{3eqm1}) of $v(\xi)$ along with (\ref{3eqm2})
and write the result in the form
\begin{equation}\label{3eqn}
Q=\frac{1}{T}\sum_{i=1}^\nu\left( K_i+F_{(1)}[1]\,L_i  \right),
\end{equation}
where $K_i, L_i, T$ are polynomials in $F$ whose degrees depend linearly on $m, \hat{m}$ such that
\begin{equation}
M_i=\frac{ 1}{T}\left( K_i+F_{(1)}[1]\,L_i \right).
\end{equation}
  Solve the system of linear algebraic equations  obtained by  balancing the degree of
$K_1$ with that of $K_2$ and  the degree of $L_1$ with that of $L_2$ in $F$ to determine the values of  integers $m, \hat{m}.$

\subsection*{Estimation of the constants $a_i,b_i,\alpha_i.$}

Introduce into  (\ref{3eqn}) the obtained values of $m, \hat{m}.$ Write the result in the form
\begin{equation}
Q=\frac{ 1 }{T}\left(K+F_{(1)}[1]\,L \right),
\end{equation}
where
$K=\sum_{i=1}^\nu K_i$ and  $  L=\sum_{i=1}^\nu L_i  .$
Set to zero all coefficients of  distinct  monomials in $K$ and $L.$ This gives a system of algebraic equations whose the
unknowns are the constants $a_i,b_i$ and $\alpha_i.$
\begin{remark}
We have the $\tanh$-expansion method if we take
$F(\xi)=\tanh(\xi),$  the $\tan$-expansion method if
$F(\xi)=\tan(\xi),$
the  $\exp$-expansion method if
$F(\xi)=\exp(\xi),$
 the $\frac{G'}{G}$-expansion method if
$F(\xi)=\frac{G'(\xi)}{G(\xi)},$ where $G''(\xi)=\alpha\, G'(\xi)+\beta \, G(\xi)$ with $\alpha, \beta \in \mathbb{R}.$
It is noticeable  that in all these cases $F'(\xi)$ is a polynomial function in $F(\xi)$ and  hence a rational function
as required.
\end{remark}

\subsection{Double function expansion method}

It aims at finding the function $v=v(\xi),$ solution of the equation (\ref{3eqs4}), into the form
\begin{equation}\label{3eqm3}
v=A(F,G)+F_{(1)}[1]B(F,G)+G_{(1)}[1]C(F,G)
+ F_{(1)}[1]G_{(1)}[1]D(F,G),
\end{equation}
where
$$  A(F,G)=\sum_{i=-m_1}^{m_1}\sum_{j=-m_2}^{m_2}a_{i,j}F^iG^j,\quad  B(F,G)=\sum_{i=-\hat{m}_1}^{\hat{m}_1}\sum_{j=-\hat{m}_2}^{\hat{m}_2}b_{i,j}F^iG^j,  $$
$$  C(F,G)=\sum_{i=-\check{m}_1}^{\check{m}_1}\sum_{j=-\check{m}_2}^{\check{m}_2}c_{i,j}F^iG^j,\quad D(F,G)=\sum_{i=-\bar{m}_1}^{\bar{m}_1}\sum_{j=-\bar{m}_2}^{\bar{m}_2}d_{i,j}F^iG^j    $$
and the functions $F=F(\xi)$ and $G=G(\xi)$  are solutions of  an  auxiliary  system of differential  equations of  the form
 \begin{eqnarray}
F_{(1)}[1]= R_1   & \emph{\emph{or}}&  \left( F_{(1)}[1]\right)^2= R_1,\label{3eqm4}\\
G_{(1)}[1]= R_2   & \emph{\emph{or}}&  \left( G_{(1)}[1]\right)^2= R_2,\label{3eqm5}
\end{eqnarray}
$R_k,$ $k=1,2$ being  rational functions in $F,$ $G$  and $m_1, m_2,\hat{m}_1, \hat{m}_2,\check{m}_1,\check{m_2},\bar{m}_1, \bar{m}_2\in \mathbb{N},$ $a_{i,j},b_{i,j},c_{i,j},d_{i,j}\in \mathbb{R}.$
The higher order derivatives of $F$ and $G$ can be written into  the form
\begin{eqnarray}
F_{(k)}[1]&=& R^{0,k}_1+ F_{(1)}[1]\,R^{1,k}_1 + G_{(1)}[1]\,R^{2,k}_1+ F_{(1)}[1]G_{(1)}[1]\,R^{3,k}_1 ,\nonumber\\
G_{(k)}[1]&=& R^{0,k}_2+ F_{(1)}[1]\,R^{1,k}_2 + G_{(1)}[1]\,R^{2,k}_2+ F_{(1)}[1]G_{(1)}[1]\,R^{3,k}_2 ,\nonumber
\end{eqnarray}
where the positive integer $k\geq 2$ and  $R^{i,k}_j,$ $i=0, 1, 2, 3,$ $j=1,2,$ are also rational functions in $F,$ $G.$

One can adopt the following  iterative process  in practice for the determination of different parameters of  $v.$
\subsection*{ Estimation of the integers $m_1, m_2,\hat{m}_1, \hat{m}_2,\check{m}_1,\check{m_2},\bar{m}_1, \bar{m}_2.$}

 Let $M_1, M_2,  \cdots  ,M_\nu$ be the monomials of $Q$ such that $M_1$ contains the highest order  derivative
of the function $v(\xi)$  and let $M_2$ be a nonlinear or linear monomial. $M_2$ is linear only in the case where all
remaining monomials are linear. Substitute in (\ref{3eqs4}) the expression (\ref{3eqm3}) of $v(\xi)$ along with (\ref{3eqm4})-(\ref{3eqm5}) and write the result in the form
\begin{equation}\label{3eqn2}
Q=\frac{1}{T}\sum_{i=1}^\nu\left( K_i+F_{(1)}[1]\,L_i +G_{(1)}[1]\,S_i +F_{(1)}[1]G_{(1)}[1]\,J_i\right),
\end{equation}
where $K_i, L_i, S_i,J_i, T$ are polynomials in $F, G$ whose the degrees linearly depend  on $m_1,$ $m_2,$ $\hat{m}_1,$ $\hat{m}_2,$ $\check{m}_1,$ $\check{m_2},$ $\bar{m}_1,$ $\bar{m}_2$ such that
\begin{equation}
M_i=\frac{ 1}{T}\left(K_i+F_{(1)}[1]\,L_i +G_{(1)}[1]\,S_i +F_{(1)}[1]G_{(1)}[1]\,J_i \right).
\end{equation}
 Solve the system of linear algebraic equations  obtained by  balancing the degree of
$K_1$ with that of $K_2,$  the degree of $L_1$ with that of $L_2,$ the degree of $S_1$ with that of $S_2$ and the degree of $J_1$ with that of $J_2$ in $F$ and $G$  to determine the values of  integers $m_1, m_2,\hat{m}_1, \hat{m}_2,\check{m}_1,\check{m_2},\bar{m}_1, \bar{m}_2.$


\subsection*{Estimation of the constants $a_{i,j},b_{i,j},c_{i,j}, d_{i,j}, \alpha_i.$}

 Introduce into  (\ref{3eqn2}) the obtained values of $m_1, m_2,\hat{m}_1, \hat{m}_2,\check{m}_1,\check{m_2},\bar{m}_1, \bar{m}_2.$ Write the result in the form
\begin{equation}
Q=\frac{ 1 }{T}\left(K+F_{(1)}[1]\,L +G_{(1)}[1]\,S +F_{(1)}[1]G_{(1)}[1]\,J \right),
\end{equation}
where
$K=\sum_{i=1}^\nu K_i  ,\quad  L=\sum_{i=1}^\nu L_i,\quad S=\sum_{i=1}^\nu S_i $ and $ J=\sum_{i=1}^\nu J_i   . $
Set to zero all coefficients of  distinct  monomials in $K,$  $L,$ $S$ and $J.$ This gives a system of algebraic equations whose the
unknowns are the constants $a_{i,j},b_{i,j},c_{i,j},d_{i,j}$ and $\alpha_i.$
\begin{remark}
We have the $(\sinh,\cosh)$-expansion method if we take
$F(\xi)=\sinh(\xi)$ and $G(\xi)=\cosh(\xi),$ the $(\sin,\cos)$-expansion method if
$F(\xi)=\sin(\xi)$ and $G(\xi)=\cos(\xi).$
We obtain a  more general formulation of the $\left(\frac{H'}{H},\frac{1}{H}\right)$-expansion method by setting
$F(\xi)=\frac{H'(\xi)}{H(\xi)}$ and $G(\xi)=\frac{1}{H(\xi)},$ where $H''(\xi)+\lambda\, H(\xi)=\mu $ with $\lambda, \mu \in \mathbb{R},$ and
a more general formulation of the projective Riccati equation expansion method if we choose $F(\xi),G(\xi)$ such that
$F'(\xi)=\alpha\,F(\xi)\,G(\xi)$    and $G'(\xi)=\mu+\alpha^2\,G^2(\xi)-\beta\,F(\xi)$ with $\alpha,\beta,\mu\in \mathbb{R}^{\star}.$
In all these cases, $F'(\xi)$ and $G'(\xi)$ are  also polynomial functions in $F(\xi),G(\xi)$ and  hence rational functions as required.
\end{remark}

\subsection{Triple function expansion method}

Here, we need to express the function $v=v(\xi),$ solution of the equation (\ref{3eqs4}), into the form
\begin{eqnarray}\label{3eqm6}
v&=&A+B^{1,1} F_{(1)}[1]
+B^{2,1} G_{(1)}[1]
+B^{3,1} H_{(1)}[1]
+C^{1,1} F_{(1)}[1]G_{(1)}[1]\nonumber\\
&+&C^{2,1} F_{(1)}[1]H_{(1)}[1]
+C^{3,1} G_{(1)}[1]H_{(1)}[1]
+D \,F_{(1)}[1]G_{(1)}[1]H_{(1)}[1]
\end{eqnarray}
where $A, B^{1,1},B^{2,1},B^{3,1},C^{1,1},C^{2,1},C^{3,1},D$ are functions of $F,G,H$ given by
\begin{eqnarray}
 A= \sum_{i=-m_1}^{m_1}\sum_{j=-m_2}^{m_2}\sum_{k=-m_3}^{m_3}a_{i,j,k}F^iG^jH^k,& & B^{1,1}=\sum_{i=-\hat{m}_{1,1}}^{\hat{m}_{1,1}}\sum_{j=-\hat{m}_{1,2}}^{\hat{m}_{1,2}}\sum_{k=-\hat{m}_{1,3}}^{\hat{m}_{1,3}}b^{1,1}_{i,j,k}F^iG^jH^k,    \nonumber\\
   B^{2,1}= \sum_{i=-\hat{m}_{2,1}}^{\hat{m}_{2,1}}\sum_{j=-\hat{m}_{2,2}}^{\hat{m}_{2,2}}\sum_{k=-\hat{m}_{2,3}}^{\hat{m}_{2,3}}b^{2,1}_{i,j,k}F^iG^jH^k ,& &  B^{3,1}=  \sum_{i=-\hat{m}_{3,1}}^{\hat{m}_{3,1}}\sum_{j=-\hat{m}_{3,2}}^{\hat{m}_{3,2}}\sum_{k=-\hat{m}_{3,3}}^{\hat{m}_{3,3}}b^{3,1}_{i,j,k}F^iG^jH^k,   \nonumber\\ C^{1,1}=\sum_{i=-\check{m}_{1,1}}^{\check{m}_{1,1}}\sum_{j=-\check{m}_{1,2}}^{\check{m}_{1,2}}\sum_{k=-\check{m}_{1,3}}^{\check{m}_{1,3}}c^{1,1}_{i,j,k}F^iG^jH^k,& &
C^{2,1}=\sum_{i=-\check{m}_{2,1}}^{\check{m}_{2,1}}\sum_{j=-\check{m}_{2,2}}^{\check{m}_{2,2}}\sum_{k=-\check{m}_{2,3}}^{\check{m}_{2,3}}c^{2,1}_{i,j,k}F^iG^jH^k , \nonumber\\
 C^{3,1}=  \sum_{i=-\check{m}_{3,1}}^{\check{m}_{3,1}}\sum_{j=-\check{m}_{3,2}}^{\check{m}_{3,2}}\sum_{k=-\check{m}_{3,3}}^{\check{m}_{3,3}}c^{3,1}_{i,j,k}F^iG^jH^k ,& &  D=\sum_{i=-\bar{m}_1}^{\bar{m}_1}\sum_{j=-\bar{m}_2}^{\bar{m}_2}\sum_{k=-\bar{m}_3}^{\bar{m}_3}d_{i,j,k}F^iG^jH^k  \nonumber
\end{eqnarray}
and $m_i, \bar{m}_i, \check{m}_{i,j}, \hat{m}_{i,j}$ are positive integers, $a_{i,j,k},b^{l,1}_{i,j,k},c^{l,1}_{i,j,k},d_{i,j,k}\in \mathbb{R};$ the functions $F=F(\xi),$ $G=G(\xi)$ and $H=H(\xi)$  are solutions of  the following  auxiliary  system of differential  equations:
 \begin{eqnarray}
F_{(1)}[1]= R_1   & \emph{\emph{or}}&  \left( F_{(1)}[1]\right)^2= R_1,\label{3eqm7}\\
G_{(1)}[1]= R_2   & \emph{\emph{or}}&  \left( G_{(1)}[1]\right)^2= R_2,\label{3eqm8}\\
H_{(1)}[1]= R_3   & \emph{\emph{or}}&  \left( H_{(1)}[1]\right)^2= R_3,\label{3eqm9}
\end{eqnarray}
$R_k,$ $k=1,2,3$ being  rational functions in $F,$ $G,$ $H.$
The higher order derivatives of $F,$ $G$ and $H$ can be written into  the form
\begin{eqnarray}
F_{(k)}[1]&=& R^{0,k}_1+ F_{(1)}[1]\,R^{1,k}_1 + G_{(1)}[1]\,R^{2,k}_1+ H_{(1)}[1]\,R^{3,k}_1+ F_{(1)}[1]G_{(1)}[1]\,R^{4,k}_1\nonumber\\
&+& F_{(1)}[1]H_{(1)}[1]\,R^{5,k}_1 + G_{(1)}[1]H_{(1)}[1]\,R^{6,k}_1+ F_{(1)}[1]G_{(1)}[1]H_{(1)}[1]\,R^{7,k}_1 ,\nonumber\\
G_{(k)}[1]&=& R^{0,k}_2+ F_{(1)}[1]\,R^{1,k}_2 + G_{(1)}[1]\,R^{2,k}_2+ H_{(1)}[1]\,R^{3,k}_2+ F_{(1)}[1]G_{(1)}[1]\,R^{4,k}_2\nonumber\\
&+& F_{(1)}[1]H_{(1)}[1]\,R^{5,k}_2 + G_{(1)}[1]H_{(1)}[1]\,R^{6,k}_2+ F_{(1)}[1]G_{(1)}[1]H_{(1)}[1]\,R^{7,k}_2 ,\nonumber\\
H_{(k)}[1]&=& R^{0,k}_3+ F_{(1)}[1]\,R^{1,k}_3 + G_{(1)}[1]\,R^{2,k}_3+ H_{(1)}[1]\,R^{3,k}_3+ F_{(1)}[1]G_{(1)}[1]\,R^{4,k}_3\nonumber\\
&+& F_{(1)}[1]H_{(1)}[1]\,R^{5,k}_3 + G_{(1)}[1]H_{(1)}[1]\,R^{6,k}_3+ F_{(1)}[1]G_{(1)}[1]H_{(1)}[1]\,R^{7,k}_3 ,\nonumber
\end{eqnarray}
where the positive integer $k\geq 2$ and  $R^{i,k}_j,$ $i=0, 1, 2, 3,4,5,6,7,$ $j=1,2,3,$ are also rational functions in $F,$ $G,$ $H.$
In practice, the following  iterative process leads to  the determination of different parameters of  $v.$

 \subsection*{Estimation of the integer $m_i, \bar{m}_i, \check{m}_{i,j}, \hat{m}_{i,j}.$}

Let $M_1, M_2,  \cdots  ,M_\nu$ be the monomials of $Q$ such that $M_1$ contains the highest order  derivative
of the function $v(\xi)$  and let $M_2$ be a nonlinear or linear monomial. $M_2$ is linear only in the case where all
remaining monomials are linear. Substitute in (\ref{3eqs4}) the expression (\ref{3eqm6}) of $v(\xi)$ along with (\ref{3eqm7})-(\ref{3eqm9}) and write  the result in the form
\begin{eqnarray}
Q&=&\frac{1}{T}\sum_{i=1}^\nu\left( K_i+ F_{(1)}[1]\,L_{1,i} + G_{(1)}[1]\,L_{2,i}+ H_{(1)}[1]\,L_{3,i}+ F_{(1)}[1]G_{(1)}[1]\,S_{1,i} \right)\nonumber\\
&+&\frac{1}{T}\sum_{i=1}^\nu\left( F_{(1)}[1]H_{(1)}[1]\,S_{2,i} + G_{(1)}[1]H_{(1)}[1]\,S_{3,i}+ F_{(1)}[1]G_{(1)}[1]H_{(1)}[1]\,J_{i} \right),\label{3eqn3}
\end{eqnarray}
where $K_i,$  $L_{j,i},$  $S_{j,i},$ $J_i,$ $T$ are polynomials in $F, G, H$ whose the degrees linearly depend  on $m_i, $ $\bar{m}_i,$ $\check{m}_{i,j},$ $\hat{m}_{i,j}$ and
\begin{eqnarray}
M_i&=&\frac{ 1 }{T}\left(K_i+ F_{(1)}[1]\,L_{1,i} + G_{(1)}[1]\,L_{2,i}+ H_{(1)}[1]\,L_{3,i}+ F_{(1)}[1]G_{(1)}[1]\,S_{1,i} \right)\nonumber\\
&+&\frac{ 1 }{T}\left(F_{(1)}[1]H_{(1)}[1]\,S_{2,i} + G_{(1)}[1]H_{(1)}[1]\,S_{3,i}+ F_{(1)}[1]G_{(1)}[1]H_{(1)}[1]\,J_{i}\right).
\end{eqnarray}
Solve the system of linear algebraic equations  obtained by  balancing the degree of
$K_1$ with that of $K_2,$ the degree of $L_{j,1}$ with that of $L_{j,2},$  the degree of $S_{j,1}$ with that of $S_{j,1}$
 and the degree of $J_1$ with that of $J_2$ in $F,$  $G$  and $H$ to determine the values of  the
 integers $m_i, \bar{m}_i, \check{m}_{i,j}, \hat{m}_{i,j}.$

 \subsection*{Estimation of the constants $a_{i,j,k},b^{l,1}_{i,j,k},c^{l,1}_{i,j,k},d_{i,j,k}, \alpha_i.$}

 Introduce into  (\ref{3eqn3}) the obtained values of $m_i, \bar{m}_i, \check{m}_{i,j}, \hat{m}_{i,j}.$ Write the result in the form
\begin{eqnarray}
Q&=&\frac{ 1 }{T}\left(  K+ F_{(1)}[1]\,L_{1} + G_{(1)}[1]\,L_{2}+ H_{(1)}[1]\,L_{3}+ F_{(1)}[1]G_{(1)}[1]\,S_{1} \right)\nonumber\\
&+&\frac{ 1 }{T}\left(F_{(1)}[1]H_{(1)}[1]\,S_{2} + G_{(1)}[1]H_{(1)}[1]\,S_{3}+ F_{(1)}[1]G_{(1)}[1]H_{(1)}[1]\,J\right).
\end{eqnarray}
where
$
K=\sum_{i=1}^\nu K_i  ,\quad
L_j=\sum_{i=1}^\nu L_{j,i} ,\quad
S_j=\sum_{i=1}^\nu S_{j,i}  $
and $
J=\sum_{i=1}^\nu J_i .
$\\
Set to zero all coefficients of  distinct  monomials in $K,$  $L_j,$  $S_j$ and
$J.$ This gives a system of algebraic equations whose the unknowns are the constants
 $a_{i,j,k},b^{l,1}_{i,j,k},c^{l,1}_{i,j,k},d_{i,j,k}$ and $\alpha_i.$
\begin{remark}
We have  the  $(\emph{sn},\emph{cn},\emph{dn})$-expansion method if we take
$F(\xi)=\emph{sn}(\xi),$ $G(\xi)=\emph{cn}(\xi)$ and $H(\xi)=\emph{dn}(\xi),$
where $\emph{sn}(\xi)=\emph{sn}(\xi,k),$ $\emph{cn}(\xi)=\emph{cn}(\xi,k)$ and $\emph{dn}(\xi)=\emph{dn}(\xi,k)$
 with $0  < k  < 1$ are the basis Jacobi elliptic functions \cite{81r}.
The function $\emph{sn},$    $\emph{cn}$  and $ \emph{dn}$ are solutions of the first order ordinary differential
equations
\begin{equation}\label{3eqs11}
\left(  w'(\xi)\right)^2= \left( 1-w^2(\xi) \right) \left( 1-k^2 w^2(\xi) \right),
\end{equation}
\begin{equation}\label{3eqs12}
\left(  w'(\xi)\right)^2= \left( 1-w^2(\xi) \right) \left( k^2 w^2(\xi) +1-k^2  \right),
\end{equation}
\begin{equation}\label{3eqs13}
\left(  w'(\xi)\right)^2= \left( 1+w^2(\xi) \right) \left(  w^2(\xi) +k^2-1  \right),
\end{equation}
respectively. When $k\rightarrow 0$ the Jacobi functions degenerate to the trigonometric functions, i.e.,
$$ \lim_{k\rightarrow 0}\emph{sn}(\xi,k)=\sin(\xi) , \quad \lim_{k\rightarrow 0}\emph{cn}(\xi,k)=\cos(\xi), \quad \lim_{k\rightarrow 0}\emph{dn}(\xi,k)=1.    $$
When $k\rightarrow 1,$ the Jacobi functions degenerate to the hyperbolic functions, i.e.,
$$ \lim_{k\rightarrow 1}\emph{sn}(\xi,k)=\tanh(\xi) , \quad \lim_{k\rightarrow 1}\emph{cn}(\xi,k)=\frac{1}{\cosh(\xi)}, \quad \lim_{k\rightarrow 1}\emph{dn}(\xi,k)=\frac{1}{\cosh(\xi)}.    $$
The function $\emph{sn},$    $\emph{cn}$  and $ \emph{dn}$ have the  algebraic  properties
$$ \emph{sn}^2(\xi,k)+\emph{cn}^2(\xi,k)=1,\quad k^2 \emph{sn}^2(\xi,k)+\emph{dn}^2(\xi,k)=1,    $$
$$ k^2 \emph{cn}^2(\xi,k)+1-k^2=\emph{dn}^2(\xi,k),\quad \emph{cn}^2(\xi,k)+\left(1-k^2\right)\emph{sn}^2(\xi,k)=\emph{dn}^2(\xi,k)   $$
and the differential properties
$$   \frac{d}{d \xi} \emph{sn}(\xi,k)=\emph{cn}(\xi,k) \,\emph{dn}(\xi,k),\quad  \frac{d}{d \xi} \emph{cn}(\xi,k)=-\emph{sn}(\xi,k)\, \emph{dn}(\xi,k),$$
$$ \frac{d}{d \xi} \emph{dn}(\xi,k)=-k^2 \emph{sn}(\xi,k) \,\emph{cn}(\xi,k) .   $$
\end{remark}

\section{Application to some relevant NLPDEs  in physics}
\begin{itemize}
\item[(A)] Case $1$: Burger-Fisher equation \label{ex1}

Let us use the single function expansion method to analyze the
Burger-Fisher equation \cite{1r}
\begin{equation}\label{3eqh1}
u_{xx}+uu_x-u_t+u-u^2=0,
\end{equation}
where $u=u(t,x).$ We first substitute into (\ref{3eqh1}) the variables $u(t,x)=v(\xi),$  $\xi=\alpha \,t+\beta\,x,$
with $\alpha,\beta\in \mathbb{R}$ to obtain
\begin{equation}\label{3eqh2}
\beta^2\,v_{\xi\xi}+\beta\,vv_\xi-\alpha\,v_\xi+v-v^2=0.
\end{equation}
We take the expansion (\ref{3eqm1}) to look for the solution of the equation (\ref{3eqh2}) with the assumption that the function $F=F(\xi)$ satisfies the auxiliary equation $F'(\xi)=1-F^2(\xi).$
Balancing the degree of the nonlinear term $vv_\xi$ with that of the linear term $v_{\xi\xi}$ yields $m=\hat{m}=1.$
This allows to take the ansatze
 \begin{equation}\label{3eqh3}
v(\xi)=\sum_{i=-1}^{1}a_iF^i(\xi)+F'(\xi)\sum_{j=-1}^{1}b_jF^j(\xi),
\end{equation}
where $a_i,b_j$ are constants to  determine. Here $F(\xi)=\tanh(\xi)$ and (\ref{3eqh3}) becomes
\begin{equation}\label{3eqh4}
v(\xi)=-b_1\tanh^3(\xi)-b_0\tanh^2(\xi)+\left(a_1+b_1-b_{-1}  \right)\tanh(\xi)+a_0+b_0+\frac{a_{-1}+b_{-1}}{\tanh(\xi)}.
\end{equation}
Substitute (\ref{3eqh4}) into (\ref{3eqh2}). Then,
collect and set to zero the coefficients of each power of $\tanh^i(\xi).$ Solving the resulting system with respect to
the parameters $a_i, b_j, \alpha, \beta,$   several solution sets are obtained  leading to the following
solutions for the equation (\ref{3eqh1}): $\quad \epsilon\in \{-1,+1\}$
\begin{eqnarray}
 u_1(t,x)&=&0,\nonumber\\
 u_2(t,x)&=&1,\nonumber\\
 u_3(t,x)&=&\frac{1}{2}+\frac{\frac{\epsilon}{2}}{\tanh\left(\frac{\epsilon}{2}t\right)},\nonumber\\
 u_4(t,x)&=& \frac{1}{2}+\frac{\epsilon}{2} \tanh\left(\frac{\epsilon}{2}t\right)     ,\nonumber\\
 u_5(t,x)&=& \frac{1}{2}+ \frac{\frac{\epsilon}{2}}{ \tanh\left[\frac{\epsilon}{4}\left(\frac{5}{2}t+x\right)\right]},\nonumber\\
 u_6(t,x)&=& \frac{1}{2}+\frac{\epsilon}{2} \tanh\left[\frac{\epsilon}{4}\left(\frac{5}{2}t+x\right)\right]    ,\nonumber\\
 u_7(t,x)&=& \frac{1}{2}+\frac{\epsilon}{4}  \tanh\left(\frac{\epsilon}{4}t\right) + \frac{\frac{\epsilon}{4}}{\tanh\left(\frac{\epsilon}{4}t\right)},\nonumber\\
 u_8(t,x)&=& \frac{1}{2}+\frac{\epsilon}{4}  \tanh\left[\frac{\epsilon}{8}\left(\frac{5}{2}t+x\right)\right] + \frac{\frac{\epsilon}{4}}{\tanh\left[\frac{\epsilon}{8}\left(\frac{5}{2}t+x\right)\right]}.\nonumber
 \end{eqnarray}

\item[(B)] Case $2$: Burger-Fisher equation

Consider now the double function expansion method for the analysis of the
Burger-Fisher equation \cite{1r}
\begin{equation}\label{3eqhh1}
u_{xx}+uu_x-u_t+u-u^2=0,
\end{equation}
where $u=u(t,x).$ We first substitute into (\ref{3eqhh1}) the variables $u(t,x)=v(\xi),$  $\xi=\alpha \,t+\beta\,x,$
with $\alpha,\beta\in \mathbb{R}$ to obtain
\begin{equation}\label{3eqhh2}
\beta^2\,v_{\xi\xi}+\beta\,vv_\xi-\alpha\,v_\xi+v-v^2=0.
\end{equation}
We take the expansion (\ref{3eqm3}) to look for the solution of the equation (\ref{3eqhh2})
assuming  that the functions $F=F(\xi)$ and $G=G(\xi)$ satisfy the auxiliary
equation $F'(\xi)=G(\xi)$ and $G'(\xi)=F(\xi).$
Balancing the degree of the nonlinear term $vv_\xi$ with that of the linear term $v_{\xi\xi}$ yields $m_1=m_2=\hat{m}_1= \hat{m}_2=\check{m}_1=\check{m_2}=\bar{m}_1= \bar{m}_2=2.$
This allows to take the ansatze
\begin{equation}\label{3eqmh3}
v=\sum_{i=-2}^{2}\sum_{j=-2}^{2}  \left\{   a_{i,j}+b_{i,j} F_{(1)}[1] +c_{i,j}G_{(1)}[1]
+ d_{i,j}F_{(1)}[1]G_{(1)}[1]\right\}F^iG^j ,
\end{equation}
where $a_{i,j},b_{i,j},c_{i,j},d_{i,j}$ are constants to  determine.
 Note that  $F(\xi)=\cosh(\xi)$ and $G(\xi)=\sinh(\xi).$
Substitute (\ref{3eqmh3}) into (\ref{3eqhh2}),
collect and set to zero the coefficients of each power of $\cosh^i(\xi)\sinh^j(\xi).$ Solving the resulting system with respect to
the parameters $a_{i,j},$ $b_{i,j},$ $c_{i,j},$ $d_{i,j},$  $ \alpha,$  $ \beta,$   several solution sets are obtained  leading, in addition to the
solutions  found  in the Case $1$, to  the following
solutions of equation (\ref{3eqhh1}), ($a$ is an arbitrary constant):
\begin{eqnarray}
 u_{9}(t,x)&=&  a \left[\sinh(2t+x)+\cosh(2t+x) \right],\nonumber\\
 u_{10}(t,x)&=&  1+a \left[\sinh(t+x)+\cosh(t+x) \right]   ,\nonumber\\
 u_{11}(t,x)&=& a \left[\sinh\left(t+\frac{1}{2}x\right)+\cosh\left(t+\frac{1}{2}x\right) \right]^2 ,\nonumber\\
 u_{12}(t,x)&=& 1+a \left[\sinh\left(\frac{1}{2}t+\frac{1}{2}x\right)+\cosh\left(\frac{1}{2}t+\frac{1}{2}x\right) \right]^2 .\nonumber
\end{eqnarray}

\item[(C)] Case $3$: Fifth-order KdV equations

Consider the well known fifth-order KdV (fKdV) equations \cite{1r} in its standard form:
\begin{equation}\label{5eqsss1}
u_t+\sigma u^2u_x +\delta u_x u_{2x} +\rho uu_{3x} +u_{5x}=0,
\end{equation}
where $\sigma, \delta, \rho$ are arbitrary nonzero  real parameters and $u=u(t,x)$ is a sufficiently smooth function.
 A variety of the fKdV equations can be retrieved from this equation by changing the real values of the parameters $\sigma, \delta$ and  $\rho.$
 However, five well known forms of the fKdV equations are of particular interest in the literature.   There are:
 \begin{itemize}
 \item[(C1)] The Sawada-Kotera (SK) equation \cite{79r}  given by
 \begin{equation}\label{5eqsss2}
u_t+5 u^2u_x +5 u_x u_{2x} +5 uu_{3x} +u_{5x}=0;
\end{equation}
\item[(C2)] The Caudrey-Dodd-Gibbon equation \cite{92r}  given by
\begin{equation}\label{5eqsss3}
u_t+180 u^2u_x +30 u_x u_{2x} +30 uu_{3x} +u_{5x}=0;
\end{equation}
\item[(C3)] The Lax equation \cite{82r} provided by
\begin{equation}\label{5eqsss4}
u_t+30 u^2u_x +20 u_x u_{2x} +10 uu_{3x} +u_{5x}=0;
\end{equation}
\item[(C4)] The Kaup-Kupersmidt (KK) equation \cite{85r} expressed as
\begin{equation}\label{5eqsss5}
u_t+20 u^2u_x +25 u_x u_{2x} +10 uu_{3x} +u_{5x}=0.
\end{equation}
\item[(C5)] The Ito equation \cite{88r} written as
\begin{equation}\label{5eqsss6}
u_t+2 u^2u_x +6 u_x u_{2x} +3 uu_{3x} +u_{5x}=0.
\end{equation}
\end{itemize}
It is important to note that there is another significant fifth-order equation that appears in the literature
in the form
\begin{equation}\label{5eqsss7}
u_t+\sigma u u_x +\delta u_{3x} -\rho u_{5x}=0,
\end{equation}
where $\sigma, \delta, \rho$ are constants. This equation is called the Kawahara equation. The standard form
of the Kawahara equation \cite{87r}  is a fifth order KdV equation of the form
\begin{equation}\label{5eqsss8}
u_t+6 u u_x + u_{3x} - u_{5x}=0,
\end{equation}
that describes a model for plasma waves, capillary-gravity water waves. The Kawahara equation appears in the theory
of shallow water with surface tension and in the theory of magneto acoustic waves in a cold collision free plasma.
In \cite{51r} we have studied the Kawahara equation (\ref{5eqsss7})  using the symmetry group analysis method.

In the new variables
$   y=\alpha t+\beta x\,\,$ and $\,\, v(y)=u(t,x)    ,$
  (\ref{5eqsss1}) is reduced to the ordinary differential equation
\begin{equation}\label{5eqsss9}
\alpha v_y+\beta^5 v_{5y}+\beta^3 \rho v v_{3y}+\beta \sigma v^2 v_y+\beta^3 \delta v_yv_{2y}=0
\end{equation}
while equation (\ref{5eqsss7}) is reduced to
\begin{equation}\label{5eqsss10}
\alpha v_y +\beta\sigma v v_y +\beta^3\delta v_{3y}-\beta^5 \rho v_{5y}=0.
\end{equation}
The single function expansion method suggests to take the ansatze
\begin{equation}\label{5eqsss11}
v(y)=\sum_{i=-2}^{2}a_i F^i(y)+F'(y)\sum_{j=-2}^2b_jF^j(y)
\end{equation}
for the equation (\ref{5eqsss9}) and the ansatze
\begin{equation}\label{5eqsss12}
v(y)=\sum_{i=-4}^{4}a_i F^i(y)+F'(y)\sum_{j=-4}^4 b_jF^j(y)
\end{equation}
for the equation (\ref{5eqsss10}), where $a_i, b_j$ are constants to determine and $F$ is a sufficiently smooth function.
Here, we assume that the function $F=F(y)$ satisfies the auxiliary equation $F'(y)=1-F^2(y),$ i.e. $F(y)=\tanh(y).$
Thus, (\ref{5eqsss11}) becomes
\begin{eqnarray}\label{5eqsss13}
v(y)&=& -b_2\tanh^4(y)-b_1\tanh^3(y)+\left( a_2-b_0+b_2 \right)\tanh^2(y)+\left(a_1-b_{-1}+b_1   \right)\tanh(y)\nonumber\\
&+& \frac{a_{-1}+b_{-1}}{\tanh(y)}+\frac{a_{-2}+b_{-2}}{\tanh^2(y)}
\end{eqnarray}
and (\ref{5eqsss12}) becomes
\begin{eqnarray}\label{5eqsss14}
v(y)&=& -b_4\tanh^6(y)-b_3\tanh^5(y)+\left(  a_4-b_2+b_4   \right)\tanh^4(y)+\left(a_3-b_1+b_3   \right)\tanh^3(y)\nonumber\\
&+& \left(a_2-b_0+b_2  \right)\tanh^2(y)+a_0-b_{-2}+b_0 +\frac{a_{-1}-b_{-3}+b_{-1}}{\tanh(y)}\nonumber\\
&+& \frac{a_{-2}-b_{-4}+b_{-2}}{\tanh^2(y)}+\frac{a_{-3}+b_{-3}}{\tanh^3(y)}+\frac{a_{-4}+b_{-4}}{\tanh^4(y)}.
\end{eqnarray}
Substitute (\ref{5eqsss13}) into (\ref{5eqsss9}) and (\ref{5eqsss14}) into (\ref{5eqsss10}). Then, collect and set to zero the coefficients
of each power of $\tanh(y).$ Solving the resulting system of algebraic equations, several sets of parameters $\{a_i,b_j,\alpha,\beta   \}$
are obtained  leading to the following results:
\begin{itemize}
 \item[(i)]
 Analytical solutions to the Kawahara equation (\ref{5eqsss8}):
 \begin{eqnarray}
 u_1(t,x)&=&a_0;\nonumber\\
 u_2(t,x)&=&a_0-\frac{35}{169}\left[\tanh^2(\varphi(t,x))-\frac{1}{2}\tanh^4(\varphi(t,x))\right];\nonumber\\
 u_3(t,x)&=&a_0-\frac{35}{169}\left[\frac{1}{\tanh^2(\varphi(t,x))}  -\frac{1}{2}\frac{1}{\tanh^4(\varphi(t,x))}\right];\nonumber\\
 u_4(t,x)&=&a_0+\frac{35}{5408}\left[\tanh^4(\psi(t,x)) +\frac{1}{\tanh^4(\psi(t,x))}\right]\nonumber\\
 &-&\frac{35}{1352}\left[\tanh^2(\psi(t,x))  +\frac{1}{\tanh^2(\psi(t,x))}\right],\nonumber
 \end{eqnarray}
 where $\varphi(t,x)=\frac{3\sqrt{13}(338a_0-23)}{4394}t-\frac{\sqrt{13}}{26}x\,$ and $\,\psi(t,x)=\frac{3\sqrt{13}(2704a_0-9)}{70304}t-\frac{\sqrt{13}}{52}x.$
 \item[(ii)]
 Analytical solutions to the Sawada-Kotera equation (\ref{5eqsss2}):
 \begin{eqnarray}
 u_1(t,x)&=&a_0;\nonumber\\
 u_2(t,x)&=& 8\beta^2 -12\beta^2 \tanh^2(\varphi(t,x));\nonumber\\
 u_3(t,x)&=& 8\beta^2 -12\beta^2\frac{1}{\tanh^2(\varphi(t,x))}  ;\nonumber\\
 u_4(t,x)&=& a_0 -6\beta^2 \tanh^2\left(\overline{\varphi}(t,x)\right);\nonumber\\
 u_5(t,x)&=& a_0-6\beta^2 \frac{1}{\tanh^2\left(\overline{\varphi}(t,x)\right)} ;\nonumber\\
 u_6(t,x)&=& 8\beta^2-12\beta^2\left[\tanh^2(\psi(t,x))  +\frac{1}{\tanh^2(\psi(t,x))}\right] ;\nonumber\\
 u_7(t,x)&=& a_0-6\beta^2\left[\tanh^2\left(\overline{\psi}(t,x)\right)  +\frac{1}{\tanh^2\left(\overline{\psi}(t,x)\right)}\right],\nonumber
 \end{eqnarray}
 where
 $$ \varphi(t,x)=16\beta^5 t-\beta x,\quad   \psi(t,x)=256\beta^5 t-\beta x,  $$
 $$ \overline{\varphi}(t,x)=\beta\left(76\beta^4+a_0^2-40\beta^2 a_0  \right )t-\beta x,\quad  \overline{\psi}(t,x)=\beta\left(16\beta^4+5a_0^2-40\beta^2 a_0   \right)t-\beta x.     $$
 \item[(iii)]
Analytical solutions to the Caudrey-Dodd-Gibbon equation (\ref{5eqsss3}):
 \begin{eqnarray}
 u_1(t,x)&=&a_0;\nonumber\\
 u_2(t,x)&=& \frac{4}{3}  \beta^2 -2\beta^2 \tanh^2(\varphi(t,x));\nonumber\\
 u_3(t,x)&=& \frac{4}{3}\beta^2 -2\beta^2\frac{1}{\tanh^2(\varphi(t,x))}  ;\nonumber\\
 u_4(t,x)&=& a_0 -\beta^2 \tanh^2\left(\psi(t,x)\right);\nonumber\\
 u_5(t,x)&=& a_0-\beta^2 \frac{1}{\tanh^2\left(\psi(t,x)\right)} ;\nonumber\\
 u_6(t,x)&=& \frac{4}{3}\beta^2-2\beta^2\left[\tanh^2\left(\overline{\varphi}(t,x)\right)  +\frac{1}{\tanh^2\left(\overline{\varphi}(t,x)\right)}\right];\nonumber\\
 u_7(t,x)&=& a_0-\beta^2\left[\tanh^2\left(\overline{\psi}(t,x)\right)  +\frac{1}{\tanh^2\left(\overline{\psi}(t,x)\right)}\right],\nonumber
 \end{eqnarray}
 where
 $$ \varphi(t,x)=16\beta^5 t-\beta x,\quad   \overline{\varphi}(t,x)=256\beta^5 t-\beta x,  $$
 $$ \psi(t,x)=4\beta\left(19\beta^4+45a_0^2-60\beta^2 a_0  \right )t-\beta x,\quad  \overline{\psi}(t,x)=4\beta\left(4\beta^4+45a_0^2-60\beta^2 a_0   \right)t-\beta x.     $$
\item[(iv)]
Analytical solutions to the Lax equation (\ref{5eqsss4}):
 \begin{eqnarray}
 u_1(t,x)&=&a_0;\nonumber\\
 u_2(t,x)&=& 4  \beta^2 -6\beta^2 \tanh^2(\varphi(t,x));\nonumber\\
 u_3(t,x)&=& 4 \beta^2 -6\beta^2\frac{1}{\tanh^2(\varphi(t,x))}  ;\nonumber\\
 u_4(t,x)&=& a_0 -2 \beta^2 \tanh^2\left(\psi(t,x)\right);\nonumber\\
 u_5(t,x)&=& a_0-2 \beta^2 \frac{1}{\tanh^2\left(\psi(t,x)\right)} ;\nonumber\\
 u_6(t,x)&=& 4\beta^2-2\beta^2\left[\tanh^2\left(\phi(t,x)\right)  +3\frac{1}{\tanh^2\left(\phi(t,x)\right)}\right];\nonumber\\
 u_7(t,x)&=& 4\beta^2-2\beta^2\left[3 \tanh^2\left(\phi(t,x)\right)  +\frac{1}{\tanh^2\left(\phi(t,x)\right)}\right];\nonumber\\
 u_8(t,x)&=& 4\beta^2-6\beta^2\left[\tanh^2\left(\overline{\varphi}(t,x)\right)  +\frac{1}{\tanh^2\left(\overline{\varphi}(t,x)\right)}\right];\nonumber\\
 u_9(t,x)&=& a_0-2\beta^2\left[\tanh^2\left(\overline{\psi}(t,x)\right)  +\frac{1}{\tanh^2\left(\overline{\psi}(t,x)\right)}\right],\nonumber
 \end{eqnarray}
 where
 $$ \varphi(t,x)=56\beta^5 t-\beta x,\quad   \overline{\varphi}(t,x)=896\beta^5 t-\beta x,\quad \phi(t,x)=336\beta^5 t-\beta x,  $$
 $$ \psi(t,x)=2\beta\left(28\beta^4+15a_0^2-40\beta^2 a_0  \right )t-\beta x,\,\,  \overline{\psi}(t,x)=2\beta\left(48\beta^4+15a_0^2-40\beta^2 a_0   \right)t-\beta x.     $$
\item[(v)]
Analytical solutions to the Kaup-Kupersmidt equation (\ref{5eqsss5}):
 \begin{eqnarray}
 u_1(t,x)&=&a_0;\nonumber\\
 u_2(t,x)&=&   \beta^2 -\frac{3}{2}\beta^2 \tanh^2(\varphi(t,x));\nonumber\\
 u_3(t,x)&=& \beta^2 - \frac{3}{2}\beta^2\frac{1}{\tanh^2(\varphi(t,x))}  ;\nonumber\\
 u_4(t,x)&=& 8 \beta^2 -12 \beta^2 \tanh^2\left(\psi(t,x)\right);\nonumber\\
 u_5(t,x)&=& 8 \beta^2- 12 \beta^2 \frac{1}{\tanh^2\left(\psi(t,x)\right)} ;\nonumber\\
 u_6(t,x)&=& \beta^2-\frac{3}{2} \beta^2\left[\tanh^2\left(\overline{\varphi}(t,x)\right)  +\frac{1}{\tanh^2\left(\overline{\varphi}(t,x)\right)}\right];\nonumber\\
 u_7(t,x)&=& 8 \beta^2- 12 \beta^2\left[\tanh^2\left(\overline{\psi}(t,x)\right)  +\frac{1}{\tanh^2\left(\overline{\psi}(t,x)\right)}\right],\nonumber
 \end{eqnarray}
 where
 $$ \varphi(t,x)=\beta^5 t-\beta x,\quad   \overline{\varphi}(t,x)=16\beta^5 t-\beta x,  \quad
  \psi(t,x)=176\beta^5 t-\beta x,\quad  \overline{\psi}(t,x)=2816\beta^5 t-\beta x.     $$
\item[(vi)]
Analytical solutions to the Ito equation (\ref{5eqsss6}):
 \begin{eqnarray}
 u_1(t,x)&=&a_0;\nonumber\\
 u_2(t,x)&=&   4\beta^2 -6\beta^2 \tanh^2(\beta x);\nonumber\\
 u_3(t,x)&=& 4 \beta^2 - 6\beta^2\frac{1}{\tanh^2(\beta x)}  ;\nonumber\\
 u_4(t,x)&=& 20 \beta^2 -30 \beta^2 \tanh^2\left(96\beta^5 t-\beta x\right);\nonumber\\
 u_5(t,x)&=& 20 \beta^2- 30 \beta^2 \frac{1}{\tanh^2\left(96\beta^5 t-\beta x\right)} ;\nonumber\\
 u_6(t,x)&=& 4 \beta^2-6 \beta^2\left[\tanh^2\left(\beta x\right)  +\frac{1}{\tanh^2\left(\beta x\right)}\right];\nonumber\\
 u_7(t,x)&=& 20 \beta^2- 30 \beta^2\left[\tanh^2\left(1536\beta^5 t-\beta x\right)  +\frac{1}{\tanh^2\left(1536\beta^5 t-\beta x\right)}\right].\nonumber
 \end{eqnarray}
\end{itemize}
\end{itemize}
$ $\\

Finally, let us emphasize that all these methods  involve cumbersome computations and hence do require the use of powerful computers. Except for this disadvantage,  our analysis can be straightforwardly extended  to multiple function
 expansion methods and to the investigation of  systems of partial differential equations.

\subsection*{Acknowledgements}
This work is partially supported by the Abdus Salam International
Centre for Theoretical
Physics (ICTP, Trieste, Italy) through the
OEA-ICMPA-Prj-15. The ICMPA is in partnership with
the Daniel Iagolnitzer Foundation (DIF), France. MNH is grateful to Professor H. V. Mweene for the hospitality offered during his stay at the University of Zambia where this work has been completed. He also acknowledges the Academy of Sciences for Developing World (TWAS) Research Professorship Programme and the University of Zambia in support of his Visiting Professorship at the University of Zambia.

\label{lastpage-01}

\begin{thebibliography}{99}

\bibitem{21r} M. A. Abdou,   \textit{Electronic Journal of Theoretical Physics}  4 No. 14,  (2007) 17-30.
\bibitem{29r} M. A.  Abdou and  A.  Elhanbaly,  \textit{Commun. Nonlinear Sci. Numer. Simul.}  12,  (2007) 1229-1241.
\bibitem{30r}  M. A. Abdou,  \textit{Journal of Nonlinear Dynamics} 52 (1-2), (2008)  95-102.
\bibitem{19r} A. Bekir, \textit{Phys. Scr.} 77, (2008).
\bibitem{22r} A. Bekir,  \textit{Phys. Lett.} A 372, (2008) 3400-3406.
\bibitem{37r} L. Bin  and Z. Hong-Qing, \textit{Commun. Theor. Phys.} (Beijing, China) 50, (2008)  814-820.
\bibitem{81r} C. A. Briot and J. C. Bouquet,  \textit{Th\'eorie des fonctions elliptiques}, Deuxi\`eme \'edition, Paris, Gauthier-Villars, 1875.
\bibitem{92r}    P. J. Caudrey, R. K. Dodd and J. D. Gibbon, \textit{Proc. Roy. Soc. Lond.} A 351,  (1976) 407-422.
\bibitem{24r} W. Deng-Shan, R. Yu-Jie  and Z. Hong-Qing,  \textit{Appl. Math. E-Notes} 5, (2005) 157-163.
\bibitem{28r} S. A. El-Wakil, E. M. Abulwafa, A. Elhanbaly and  M. A. Abdou, \textit{Chaos Solitons and  Fractals} 33, (2007) 1512-1522.
\bibitem{33r} A. Elgarayhi and A. A. Karawia,  \textit{Int. J. Nonlinear Sci.}  Vol.7 No.4,  (2009) 414-419.
\bibitem{16r} J. H. He and X. H. Wu,  \textit{Chaos Solitons and Fractals} v30, (2006) pp. 700-708.
\bibitem{51r} M. N. Hounkonnou and P. A. Dkengne Sielenou, \textit{Int. J. Contemp. Math. Sci.}  \textbf{4} (35), (2009) pp 1719-1738.
 \bibitem{hk1} M. N. Hounkonnou and P. A. Dkengne Sielenou,
 XXVIII Workshop  on  Geometrical  Methods  in  Physics. AIP Conference Proceedings, Volume 1191,  (2009) pp. 104-109.

 \bibitem{hk2} M.  N.  Hounkonnou and P.  A.  Dkengne  Sielenou,
 XXIX Workshop on   Geometric  Methods    in   Physics. AIP Conference Proceedings, Volume 1307,  (2010) pp. 83-88.


\bibitem{hk3}  M. N. Hounkonnou and P. A. Dkengne Sielenou, Commun. Math. Anal.,
{\it Special Volume in Honor of Prof. Peter Lax} Volume 8 Number 3,  (2010) pp. 102-119.

\bibitem{hk4}  M. N. Hounkonnou and P. A. Dkengne Sielenou,  Commun. Math. Anal.,
{\it Special Volume in Honor of Prof. Stephen Smale} Volume 10, Number 1, (2011) pp. 53-74.

\bibitem{hk5}  M. N. Hounkonnou and P. A. Dkengne Sielenou, Afr. Diaspora J. Math.,
 \textit{Special Volume in Honor of Profs. C. Corduneanu, A. Fink, and S. Zaidman} Volume 12, Number 2, (2011) pp. 73-103.
\bibitem{88r}  M. Ito, \textit{J. Phys. Soc. Japan} 49 (2), (1980) 771-778.
\bibitem{87r}  T. Kawahara, \textit{J. Phys. Soc. Japan} 33, (1972) 260-264.

\bibitem{25r} H. Kheiri, A. Jabbari,  \textit{Acta universitatis apulensis} 22, (2010) 185-194.
\bibitem{85r}  B. A. Kuperschmidt, \textit{Phys. Lett.} A 102, (1984) 213-215.
\bibitem{82r}  P. D. Lax, \textit{Commun. Pure Appl. Math.}, 21,  (1968) 467-490.
\bibitem{20r} B. Li and C. Jinlan,  \textit{Journal of Henan University of Science and Technology} (Natural Science)
26 (5), (2005) 80-83.
\bibitem{23r} L. Ling-xiao,  L. Er-qiang and  W. Ming-liang,
\textit{Appl. Math. J. Chinese Univ.} 25 (4), (2010) 454-462.
\bibitem{18r} S. Liu, Z. Fu and  Q. Zhao, \textit{Phys. Lett.} A. 289 (1), (2001) 69-74.
\bibitem{34r} J. B. Liu and  K. Q. Yang,  \textit{Chaos Solitons Fractals} 22 (1), (2004) 111-121.
\bibitem{17r} W. Malfliet and W. Hereman,  \textit{Phys. Scr.} 54 (1996).
\bibitem{79r}  K. Sawada and T. Kotera, \textit{Progr. Theoret. Phys.} 51,  (1974) pp. 1355-1367.
\bibitem{26r} A. M. Wazwaz,  \textit{Appl. Math. Comput.} 187, (2007) pp. 1131-1142.
\bibitem{31r} A. M. Wazwaz,  \textit{Appl. Math. Comput.} 188, (2007) pp. 1467-1475.
\bibitem{1r} A. M. Wazwaz, \textit{Partial differential equations and solitary wave theory}, Springer, 2009.
\bibitem{32r} M. Wen-Xiu, H. Tingwen  and Y. Zhang, \textit{Phys. Scr.} 82 (2010).
\bibitem{35r} S. Zhang, \textit{Chaos Solitons Fractals} 32 (4), (2007) pp. 1375-1383.
\bibitem{36r} S. Zhang,  \textit{Appl. Math. Comput.} 189 (1), (2007)  pp. 836-843.
\bibitem{27r} J. Zhang, X. L. Wei and Y. J., \textit{Phys. Lett.} A 372, (2008) pp. 3653-3658.




\end{thebibliography}
\end{document}